\documentclass[aps,prl,reprint]{revtex4-1}

\usepackage{graphicx}
\usepackage{color}
\usepackage{amsmath}
\usepackage{amssymb}
\usepackage{bm}
\usepackage{natbib}

\begin{document}

\title{Dynamical Instability in Boolean Networks as a Percolation Problem}
\author{Shane Squires}
\email{squires@umd.edu}
\author{Edward Ott}
\author{Michelle Girvan}
\affiliation{University of Maryland---College Park}
\date{\today}

\begin{abstract}
Boolean networks, widely used to model gene regulation, exhibit a phase transition between 
regimes in which small perturbations either die out or grow exponentially.  We show and 
numerically verify that this phase transition in the dynamics can be mapped onto a static 
percolation problem which predicts the long-time average Hamming distance between perturbed 
and unperturbed orbits.  
\end{abstract}

\maketitle

Boolean networks have been a prominent tool for modeling gene regulation since their introduction 
by Kauffman in 1969 \cite{kauffman69,de_jong02}.  In a Boolean network, each node is assigned a 
state, 0 or 1, which is synchronously updated at discrete time steps according to a pre-assigned 
update function which depends on the states of that node's inputs on the previous time step.  When 
used to model gene regulatory networks, each node represents a gene, and the state of the node 
indicates whether or not the gene is being expressed.  Kauffman's original considered random networks
and update functions in which each of the $N$ nodes has $K$ input links from randomly chosen nodes 
(the $N$-$K$ model).  Kauffman found numerically that when the in-degree $K$ crosses a critical 
value, there is a transition between a stable phase, in which small perturbations die out, to an 
unstable phase, in which small perturbations grow and become macroscopic.

A derivation of the critical in-degree was given by Derrida and Pomeau for annealed $N$-$K$ networks
\cite{derrida86}.  Here ``annealed'' means that the network edges and update functions are randomly 
redrawn between time steps.  They hypothesized that for large networks the stability properties of 
the annealed system are similar to those of the original frozen (non-annealed) system.  This 
hypothesis is well-supported by numerical experiments \cite{derrida86,bastolla96}, and we refer to 
it as the ``annealed approximation.''  Recent work \cite{pomerance09} has extended this approach by 
using a partial randomization, in which only the update functions (but not the network topology) are 
randomly generated at each time step.  In contrast with the annealed approximation, this 
``semi-annealed'' approximation describes the dynamics on a fixed network which may have nontrivial 
topological features such as edge assortativity \cite{maslov02}, motifs \cite{milo02}, and community 
structure \cite{cui07}.  The only necessary assumption is that the network is locally treelike (it
cannot have many short loops) \footnote{The locally treelike approximation is discussed in detail 
in \cite{pomerance09} and \cite{restrepo08}.  Configuration-model random networks with finite 
average degree are locally treelike as $N\to\infty$ \cite{newman01}.  It is quite common for 
treelike approximations to give excellent results even when the underlying network has significant 
clustering \cite{melnik11}; this was observed for Boolean networks in \cite{pomerance09}.}.

Some recent papers have derived stability properties of Boolean networks without annealing 
\cite{mozeika11,seshadhri11}.  These papers are complementary to ours in the following sense.  
Although rigorous, their results only apply to the ensemble average of random networks with 
restrictions on their network topology and/or update functions.  In contrast, because our results 
rely on the semi-annealed approximation, they can model the dynamics of a specific network.

Here, using our semi-annealed approach, we map the {\em dynamical} problem of stability on a Boolean 
network onto the {\em static} problem of network percolation in the $N\to\infty$ limit.  Previous 
authors have discussed the percolation properties of the ``frozen component'' of $N$-$K$ networks
\cite{flyvbjerg88,mihaljev06,samuelsson06}, and others have used percolation to discuss the stability
of $N$-$K$ lattices \cite{hansen88,obukhov89}.  In contrast, we show that a dynamic quantity, the 
long-time average Hamming distance between two initially close trajectories on a Boolean network, can 
be mapped onto the size of the giant out-component in a percolation problem.  We will illustrate this 
map in three different contexts.  First, we consider the well-known annealed approximation and map it 
onto percolation in the configuration model \cite{newman01}.  Second, we give a similar map from the 
semi-annealed approximation \cite{pomerance09} to weighted site percolation \cite{restrepo08}.  Finally, 
we treat a more general class of update functions by mapping to a correlated bond percolation problem.

{\em Model:} A Boolean network is a directed network of $N$ nodes, in which each node $i$ is 
assigned a state, $x_i(t)=0$ or $x_i(t)=1$, at each discrete time step $t$.  We denote the
in- and out-degrees of node $i$ by $d^\text{in}_i$ and $d^\text{out}_i$ and the set of inputs 
to node $i$ by ${\mathcal J}_i$.  A Boolean function or ``truth table'' $F_i$, fixed in time,
updates the state of each node $i$ at each time step $t$, 
$x_i(t)=F_i\left(\left\{x_j(t-1):j\in{\mathcal J}_i\right\}\right)$.

In the literature, the truth tables $F_i$ are usually generated randomly (e.g., \cite{derrida86}). 
For each combination of input states to node $i$, the value of $F_i$ is assigned to be $1$ 
with probability $p$ or $0$ with probability $1-p$, where $p$ is the ``bias probability.''  
Below, as in \cite{pomerance09}, we will consider the more general case where each $F_i$ is 
generated with a different bias $p_i$ assigned to each node $i$.  Later, we will also consider 
the case of ``canalizing'' functions, in which one input acts as a master switch for the truth 
table.  That is, input $j$ to node $i$ is canalizing if there is a state of $x_j$ which 
completely determines the value of $F_i$ independent of the other inputs to $i$. (When $x_j$ 
is not equal to its canalizing value, $F_i$ depends on the states of its other inputs.)  
Canalizing functions are thought to be common in real gene networks \cite{harris02,kauffman03}.

Consider two trajectories, $\bm{x}(t)$ and $\bm{\tilde{x}}(t)$, which evolve on the same Boolean 
network.  The initial conditions $\bm{x}(0)$ and $\bm{\tilde{x}}(0)$ differ only on a small 
randomly chosen fraction $\varepsilon$ of nodes.  We say that a node $i$ is ``damaged'' at time 
$t$ if $x_i(t)\ne\tilde{x}_i(t)$, and our goal is to predict the extent of the damage at long 
times.  Let $y_i$ be the fraction of time that node $i$ is damaged on an orbit of length $T$ as
$T\to\infty$.  The normalized long-time average Hamming distance $Y=\left\langle{y_i}\right\rangle$, 
$0\le{Y}\le1$, is used as the order parameter for the stability phase transition.  The average 
$\langle\cdot\rangle$ is taken over all nodes $i$, then over all initial conditions which differ 
on a fraction $\varepsilon$ of the nodes.

{\em Analytic Results:} First we treat the annealed approximation for random networks \cite{luque97}.  
We assume that the truth tables are randomly generated with a bias which depends only on degree.  
Let $P_{jk}$ be the probability that a node has $j$ inputs and $k$ outputs, and let the bias of 
such a node be $p_{jk}$.  We define the sensitivity \cite{shmulevich04} to be 
$q_{jk}=2p_{jk}\left(1-p_{jk}\right)\in[0,1]$, which can be interpreted as the probability that a 
node with $j$ inputs and $k$ outputs will become damaged at time $t$ if at least one of its inputs 
is damaged at time $t-1$.  

In the annealed approximation, $Y$ can be predicted analytically using a method derived in
\cite{derrida86} and \cite{lee07}, which can be explained as follows.  Let $z$ denote the average 
degree of the network, i.e.\ $z=\sum_{j,k}jP_{jk}=\sum_{j,k}k{P_{jk}}$, and let $E$ denote the 
probability that a randomly selected edge originates from a damaged node.  A randomly selected 
edge originates from a node with $j$ inputs and $k$ outputs with probability $\tfrac{kP_{jk}}{z}$, 
and such a node will become damaged with probability $q_{jk}$ if it has at least one damaged input, 
which occurs with probability $1-(1-E)^j$.  Therefore,
\begin{equation}
\begin{split}
\label{lee-Y}
E&=\sum_{j,k}\frac{kP_{jk}}{z}q_{jk}\left[1-\left(1-E\right)^j\right],\\
Y&=\sum_{j,k}P_{jk}q_{jk}\left[1-\left(1-E\right)^j\right].
\end{split}
\end{equation}
In the stable regime, these equations only have the trivial solution $E=0$ and $Y=0$, but there 
will be a nonzero solution in the unstable regime \cite{lee07}.

We now show that Eq.\ (\ref{lee-Y}) can be mapped onto the generating function formalism for treating 
weighted site percolation in directed configuration-model networks, as developed in \cite{newman01} 
and \cite{callaway00}.  In this model, each node is deleted with some probability which depends only 
on its degree.  The resulting ensemble of site-deleted networks exhibits a percolation phase transition,
above which there is a macroscopic connected component or ``giant component.''  This giant component 
contains a core of mutually path-connected nodes called the giant strongly connected component (GSCC); 
this, along with all the nodes which can be reached from it, is called the giant out-component (GOUT).  
In our map, we will identify the probability that a node is {\em not} deleted with the sensitivity, 
writing $q_{jk}$ for the probability that a node with $j$ inputs and $k$ outputs is undeleted.  With 
this identification, we will show that $Y$ maps onto the expected fraction of nodes in GOUT, which we 
denote $S$.

It is shown in \cite{newman01} and \cite{callaway00} that $S$ can be found as follows.  First, define 
generating functions for the in-degrees of nodes and edges, $F_0(w)=\sum_{j,k}P_{jk}q_{jk}w^j$ and 
$F_1(w)=\sum_{j,k}\frac{kP_{jk}}{z}q_{jk}w^j$.  Next, let $u$ be the probability that a randomly 
selected edge is not in GOUT.  The authors show through diagrammatic expansion that
\begin{equation}
\begin{split}
\label{callaway-S}
u&=1-F_1(1)+F_1(u), \\
S&=F_0(1)-F_0(u).
\end{split}
\end{equation}
We note that the substitutions $E=1-u$ and $Y=S$ map Eq.\ (\ref{lee-Y}) onto Eq.\ (\ref{callaway-S}).  
Therefore, the phase transition between dynamical stability and instability in this ensemble of random 
Boolean networks is equivalent to the static percolation phase transition on the same ensemble.  

Our second result is a more general derivation of the same correspondence, using the framework of 
\cite{pomerance09}.  This framework applies to a {\em specific} locally treelike network in which each 
node $i$ can have its own arbitrarily chosen bias $p_i$, with an associated sensitivity 
$q_i=2p_i(1-p_i)$.  Again, we will identify the sensitivity $q_i$ with a site nondeletion probability 
and map the Hamming distance, $Y$, onto the size of GOUT, $S$.  We begin by writing an analogue of Eq.\ 
(\ref{lee-Y}) for a specific node in a semi-annealed, locally treelike Boolean network,
\begin{equation}
\label{andrew-y}
y_i=q_i{\Bigg[}1-\prod_{j\in{\mathcal J}_i}\left(1-y_j\right){\Bigg]}.
\end{equation}
This is the long-time limit of a damage-spreading equation derived in \cite{pomerance09}, which noted 
that $i$ will become damaged with probability $q_i$ if at least one of its inputs is damaged.  The 
assumption that the network is locally treelike is necessary because all the probabilities in the 
product are treated as independent.

Reference \cite{restrepo08} derives a similar condition for site percolation on locally treelike 
directed networks in which the probability that each node is {\em not} deleted is $q_i$.  It defines 
$\eta_i$ as the fraction of site-deleted networks for which node $i$ is {\em not} in GOUT, and it 
shows that 
\begin{equation}
\label{roh-eta}
\eta_i=1-q_i+q_i\prod_{j\in{\mathcal J}_i}\eta_j,
\end{equation}
because a node is not in GOUT when it is either deleted or has no inputs from GOUT.  We note that 
substituting $y_i=1-\eta_i$ maps Eq.\ (\ref{andrew-y}) onto Eq.\ (\ref{roh-eta}).  Because 
$Y=\left\langle{y_i}\right\rangle$ and $S=\left\langle{1-\eta_i}\right\rangle$, this map also yields 
$Y=S$.  For $S$, the average $\langle\cdot\rangle$ is first taken over all nodes $i$, then over all 
node deletion trials.

\begin{figure}
\includegraphics[width=3.25in]{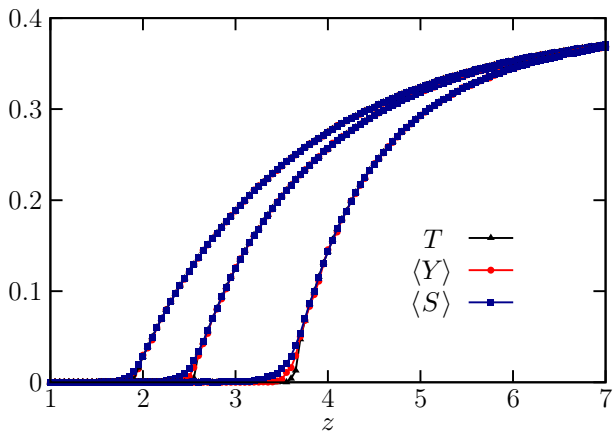}
\renewcommand\thefigure{1 (color online)}
\caption{The ensemble averages of $Y$, $S$, and $T$ (taken over $20$ networks) versus the average degree 
$z$, for three families of networks.  The three families of networks are assortative (left), neutral 
(middle), and disassortative (right).}
\end{figure}

We now introduce a third case, in which we consider Boolean networks with canalizing functions.  The 
method used for our previous results can be extended to canalizing functions, but because the truth 
table elements in a canalizing function are not generated independently, we need to consider a new 
type of percolation problem which we call correlated bond percolation.  Instead of typical bond 
percolation, in which each bond is occupied or deleted independently, we consider joint probabilities 
where the deletion of two bonds may be correlated if they are inputs to the same node.  

Here we describe a correlated bond percolation problem that corresponds to a Boolean network whose 
truth tables each have one canalizing input but are otherwise generated randomly.  That is, for each 
node $i$, there is a canalizing input $c_i$, and all the rows of the truth table on which $x_{c_i}$ 
assumes its canalizing value have the same constant output; but the outputs of the other rows are 
randomly generated with a probability bias $p_i$.  To begin, we imagine that the system is equally 
likely to be in any of its states.  As we will show, it is then formally possible to obtain equations 
describing damage spreading in closed form.  Based on our numerical results, we conjecture that these 
equations can be used to predict damage spreading in a large class of Boolean networks with frozen 
truth tables.

Working under the supposition that all system states are equally probable, we now derive an expression 
for $y_i$.  Let $r_i$ denote the ``activity'' of $c_i$ on $i$ \cite{shmulevich04}, defined as the 
fraction of states in which $i$ will become damaged if $c_i$ becomes damaged.  If $c_i$ is not damaged, 
it may be in either the canalizing or non-canalizing state, each with probability $\tfrac{1}{2}$.  In 
the first case it is impossible for $i$ to become damaged, while the second case is equivalent to Eq.\ 
(\ref{andrew-y}).  Therefore, 
\begin{equation}
\label{c-y}
y_i=r_iy_{c_i}+\frac{1}{2}q_i\left(1-y_{c_i}\right)\left[1-\prod_{j\in{\mathcal J}'_i}\left(1-y_j\right)\right],
\end{equation}
where ${\mathcal J}'_i={\mathcal J}_i-\{c_i\}$ and $q_i$ is the sensitivity of the half of the truth 
table where $x_{c_i}$ is not in its canalizing state.  It can be shown that this is equivalent to
\begin{equation}
\label{c-s}
\eta_i=1-r_i+\left(r_i-\frac{1}{2}q_i\right)\eta_{c_i}+\frac{1}{2}q_i\prod_{j\in{\mathcal J}_i}\eta_j,
\end{equation}
where $\eta_i=1-y_i$.  This corresponds to a correlated bond percolation problem in which one of the 
following three things may occur.  With probability $1-r_i$, all edges to $i$ are deleted; with 
probability $r_i-\tfrac{1}{2}q_i$, all of $i$'s edges are deleted except for the edge from $c_i$; and 
otherwise no input edges are deleted.  Note that it is straightforward to describe the case where only 
some of the nodes have a canalizing input by using Eqs.\ (\ref{c-y}-\ref{c-s}) for those nodes and 
Eqs.\ (\ref{andrew-y}-\ref{roh-eta}) for the others.  

\begin{figure}
\includegraphics[width=3.25in]{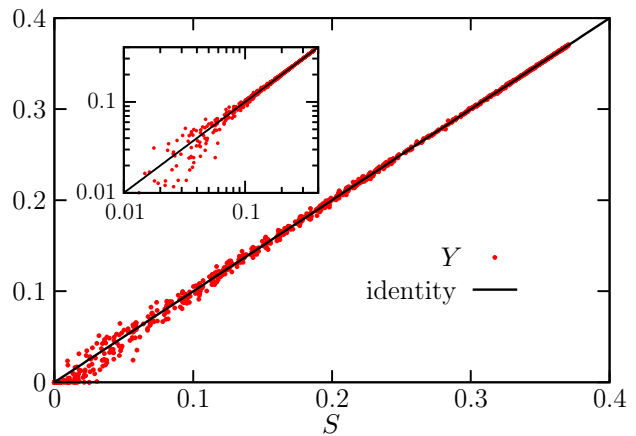}
\renewcommand\thefigure{2 (color online)}
\caption{$Y$ versus $S$ for individual neutrally assortative networks.}
\end{figure}

{\em Numerical Results:}
We begin with the map described by Eqs.\ (\ref{andrew-y}-\ref{roh-eta}), since it is more general than 
Eqs.\ (\ref{lee-Y}-\ref{callaway-S}).  We compare the long-time average Hamming distance $Y$ to the size 
of the giant out-component $S$ for particular networks.  We also compare both $Y$ and $S$ to the 
theoretical prediction given by the solution to Eq.\ (\ref{andrew-y}), which we denote $T$.

\begin{figure}
\includegraphics[width=3.366in]{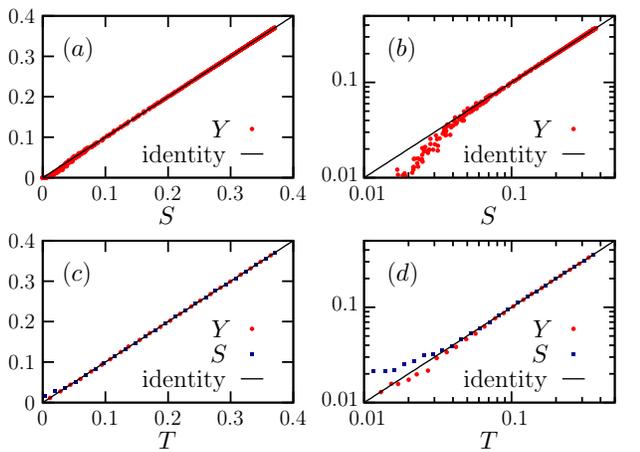}
\renewcommand\thefigure{3 (color online)}
\caption{($a$) Linear and ($b$) log-log scatterplots of $Y$ versus $S$ for data generated in the same 
way as that of Fig.\ 2, except that now we average over the quenched disorder in the truth tables as 
described in the text.  ($c$) Linear and ($d$) log-log scatterplots of $Y$ and $S$ versus $T$ for the 
same data, sampling alternate points for visibility.}
\end{figure}

Our algorithm is as follows.  First we create a configuration-model network with $N=10^5$ nodes.  The 
data in the figures were obtained using networks with Poisson-distributed in-degrees and scale-free 
out-degrees; we have also tested other degree distributions and found similar results.  If desired, 
we then enhance interesting topological features such as assortativity or feedforward loops using the 
same algorithms as in \cite{pomerance09}.  Next, we assign each node a bias $p_i$.  These may be 
distributed randomly, or, if we wish to encourage (impede) instability on the network, we distribute 
them so that the nodal average $\langle{q_i}{d_i^\text{in}}{d_i^\text{out}}\rangle$ is maximized 
(minimized) \cite{pomerance09}.  For the data in the figures, the biases $p_i$ were distributed 
randomly so that the sensitivities $q_i$ form a uniform distribution on the interval $[.3,.5]$.  We 
choose random initial conditions for $\bm{x}$, and a randomly selected fraction $\varepsilon=.01$ of 
the nodes are flipped for the initial conditions of $\bm{\tilde{x}}$.  

To find $Y$, we time-evolve the system and average $|x_i(t)-\tilde{x}_i(t)|$ between $t=900$ and 
$t=1000$, averaging over $100$ initial conditions.  The theoretical prediction is found by iterating 
Eq.\ (\ref{andrew-y}) until it converges to a solution $\bm{\hat{y}}$, then taking 
$T=\left\langle\hat{y}_i\right\rangle$.  Finding $S$ is less straightforward, because a typical 
percolation problem is only guaranteed to have a single, well-defined giant out-component in the 
$N\to\infty$ limit.  For reasons discussed in the online Supplemental Material, we choose the following 
procedure.  We delete each node $i$ with probability $1-q_i$ and find any strongly connected components 
(SCCs) in the resulting network, where we define an SCC to be a mutually path-connected set of nodes 
containing at least one loop.  We define $S$ to be the fraction of nodes which can be reached from at 
least one SCC, averaged over the ensemble of deletion trials.  We average $10^3$ deletion trials per 
network.  We find that the numerical uncertainty in our measured values of $T$, $Y$, and $S$ for each 
point in Figs.\ 1--4 is smaller than the point size; see the Supplemental Material for details.

Figure 1 illustrates the relationship between $Y$, $S$, and $T$ for networks generated in this way.  We 
see that $Y$ and $S$ have the same average values on the ensemble of random networks with given average 
degree $z$.  However, in Fig.\ 2, we see that the prediction $Y=S$ sometimes fails for individual 
networks, especially near the phase transition.  The deviations in Fig.\ 2 are primarily caused by the 
quenched disorder in the truth tables, which may cause orbits to fall onto attractors which visit 
only a small fraction of the state space (and so may deviate from the semi-annealed approximation).

\begin{figure}
\includegraphics[width=3.366in]{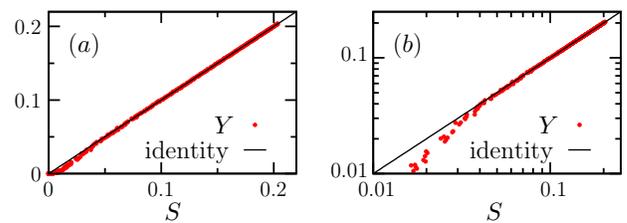}
\renewcommand\thefigure{4 (color online)}
\caption{Scatterplots of $Y$ versus $S$ for networks in which each node has one canalizing input, using 
Eqs.\ (\ref{c-y}-\ref{c-s}).}
\end{figure}

In Fig.\ 3, we have averaged over this quenched disorder by choosing truth tables from an ensemble of 
closely related frozen truth tables (but not networks) as follows.  Before we time-evolve each new 
pair of initial conditions, we perform a set of exchanges on the truth tables.  For each edge $j \to i$, 
with probability $\tfrac{1}{2}$, we exchange $x_j=0$ and $x_j=1$ on the truth table for $i$.  We note 
that there are two major differences between this and the semi-annealed approximation.  In the latter, 
the truth tables are changed {\em during} the dynamics, whereas here they are only changed before each 
new dynamical trial.  Second, whereas the semi-annealed approximation treats all inputs 
interchangeably, this procedure preserves input-specific information (such as whether an input is 
canalizing).  In Fig.\ 3, we see that this procedure yields excellent agreement between $Y$, $S$, and 
$T$ for individual networks well above the transition.  Near the transition and below it, finite-size 
effects still cause $S$ (and, to a lesser extent, $Y$) to deviate slightly from the prediction $T$.  
These effects are discussed in the Supplemental Material.

In Fig.\ 4, we perform the same numerical experiment for the case in which each node has one canalizing 
input.  We find that $Y$, $S$, and $T$ agree for individual networks when we use the map between Eqs.\ 
(\ref{c-y}) and (\ref{c-s}), but the map between Eqs.\ (\ref{andrew-y}) and (\ref{roh-eta}) fails for 
this case, indicating that we retain significant input-specific information about the dynamics when we 
average over the quenched disorder in the truth tables.

{\em Discussion:}
We have presented evidence that the stability of a Boolean network can be understood in terms of a 
related percolation problem on that network.  This relationship may be helpful in understanding the 
stability of systems modeled by Boolean networks, such as gene regulatory networks and neural networks.  
Two previously-studied cases (the annealed and semi-annealed approximations) map onto known results for 
percolation, and a case of biological interest (canalizing truth tables) maps onto a novel percolation 
problem.  These maps are valid for the typical cases in the literature (large, locally treelike 
networks with random or canalizing truth tables), but have the advantage of applying to specific 
networks rather than ensembles of random networks.  Numerical experiments show excellent agreement 
with our analysis when averaged over a family of quenched truth tables.  

{\em Acknowledgements:}
This work was funded by ONR grant N000140710734 and ARO grant W911NF1210101.

%\bibliography{bib}

\end{document}